\documentclass[12pt]{iopart}

\usepackage{iopams}  

\usepackage{graphicx,color}
\usepackage{bm}

\begin{document}
\title{Moving Walls and Geometric Phases}
\author{Paolo Facchi$^{1,2}$, Giancarlo Garnero$^{1,2}$, Giuseppe Marmo$^{3,4}$, Joseph Samuel$^{5}$}
\address{$^{1}$Dipartimento di Fisica and MECENAS, Universit\`a di Bari, I-70125  Bari, Italy}
\address{$^{2}$INFN, Sezione di Bari, I-70126 Bari, Italy}
\address{$^{3}$Dipartimento di Scienze Fisiche and MECENAS, Universit\`a di Napoli ``Federico II", I-80126  Napoli, Italy}
\address{$^{4}$INFN, Sezione di Napoli, I-80126  Napoli, Italy}
\address{$^{5}$Raman Research Institute, 560080 Bangalore, India}
\date{\today}

\begin{abstract}
We unveil the existence of a non-trivial Berry phase associated to the dynamics of a quantum particle in a one dimensional box with moving walls.
It is shown that a suitable choice of boundary conditions has to 
be made in order to preserve unitarity. For these boundary
conditions we compute explicitly the geometric phase two-form on the parameter space.
The unboundedness of the Hamiltonian describing the 
system leads to a natural prescription of renormalization for 
divergent contributions arising from the boundary.
\end{abstract}


%Keywords required only for MST, PB, PMB, PM, JOA, JOB? 
\vspace{2pc}
\noindent{\it Keywords}:  geometric phases, quantum boundary conditions, time-dependent systems
%\maketitle

\section{Introduction} \label{sec-dpw}

The case of a non-relativistic quantum particle confined in a 
one dimensional box with moving walls subject to Dirichlet boundary conditions has been investigated in great detail 
in \cite{Sara}. In this paper we consider more general boundary conditions and study the geometric phases that emerge.
The boundary conditions we focus on are those consistent with the unitarity of the dynamics as well as with 
dilation symmetry.

Geometric phases were investigated by Berry and Wilkinson~\cite{wilkinson}
who considered the behaviour of the eigenfunctions of the Laplacian in a 
region with a triangular boundary with Dirichlet boundary conditions, 
when the shape of the region was varied adiabatically. This study revealed the existence
of ``diabolical points'', shapes which have an accidental degeneracy in the spectrum. Varying 
the shape of the region in a small circuit around the diabolical point led to a reversal in the sign of the eigenfunction. Similar effects
were also noticed earlier in molecular physics~\cite{longuethiggins} as explained in the 
book by Shapere and Wilczek~\cite{shapere}. 
These sign reversals were an early example of a geometric phase.
In these problems the geometric phase is essentially of topological origin. This is because of 
the time reversal symmetry of the problem, wave functions can be chosen real and this 
constrains all geometric phases to be $1$ or $-1$. 
In a later work by Berry, the time reversal symmetry was broken by the introduction of magnetic fields and this led to 
the discovery of the full geometric phase~\cite{berryphase}, which has been subsequently studied and generalized in many directions~\cite{AA,Sam}
and widely applied~\cite{Bohm,Chrus}.

In this paper we consider a particle in a box subject to general boundary conditions, 
which (apart from some special cases) 
violate time reversal symmetry. Our Hamiltonian 
operator is the Laplacian. The location of the boundaries is adiabatically 
varied by translations and dilations,
which gives us a two parameter space of variations. 
We find that there is a geometric phase and compute the two-form on the  parameter space.
It turns out that this two-form is similar to the area two-form on the Poincar{\'e} upper half-plane. 

In section~\ref{BoundCond} we describe the most general
boundary conditions that make the Laplacian self-adjoint and we focus on a subset of these which are invariant under dilations.
In section~\ref{sec:movfix} we show how one can reduce the problem of moving walls into a fixed domain. In section~\ref{sec:Berry}, we compute the geometric phase two-form,
which measures the extent of anholonomy in a closed circuit. This calculation involves some subtleties which require a renormalization scheme.
Section~\ref{sec:Berry} gives an alternative perspective on the renormalization. Section~\ref{sec:degen} deals with the two special cases in which the boundary conditions
do not break time reversal symmetry and section~\ref{sec:conclusion} is a concluding discussion.

\section{Moving walls and quantum boundary conditions} \label{BoundCond}
In this section we use the powerful technique of \textit{boundary triples}~\cite{Brun} (see~\ref{appendix}) to classify the self-adjoint extensions of the Laplacian on an interval. We are going to use this  approach in order to find all possible boundary conditions which preserve unitarity and are invariant under dilation.

Let us consider a quantum spinless particle of mass $m$ confined in a one dimensional box $I = [a,b]$. The Hamiltonian, describing the kinetic energy of the particle, is ($\hbar=1$)
\begin{equation}
\label{eq:ham}
H \psi = \frac{p^2}{2m}\psi= -\frac{1}{2m}\psi'', \qquad \psi\in\mathrm{D}(H)= \mathcal{D} (\mathring{I}),
\end{equation}
where $\mathcal{D} (\mathring{I})$ is the space of test functions, i.e.\ the infinitely differentiable functions with compact  support in $\mathring{I}=(a,b)$.
The adjoint operator, $H^\dag$, has the same functional form of the operator~(\ref{eq:ham}) but acts on a larger space, namely  $\mathrm{D}(H^\dag)=\mathcal{H}^2(I)$, the space of square integrable functions on $I=[a,b]$ whose first and second (distribution)
derivatives are square integrable.
This Hamiltonian operator is symmetric but certainly not self-adjoint, $H\neq H^\dag$, and thus it cannot be associated to an observable of the physical system. 

The Hamiltonian's deficiency indices, determined by the equation
\begin{equation}
(H^\dag\pm \rmi\, \mathbb{I})\psi=0,
\end{equation}
are equal to 2, so that, by von Neumann's theorem (see for example \cite{Reed}), the self-adjoint extensions of the operator~(\ref{eq:ham}) are in a 
one-to-one correspondence with the unitary operators on $\mathbb{C}^2$. Unfortunately, this is a  non-constructive theorem and one needs to find other ways of working with self-adjoint extensions. 
With this end in view, we define the following maps from $D(H^\dag)$ to the space of boundary data $\mathbb{C}^2$ (see~\ref{appendix}):
\begin{eqnarray}
\label{triple}
& & \rho_1: \mathrm{D}(H^\dag)\to\mathbb{C}^2:\psi\mapsto 
\pmatrix{
\psi(a)-\rmi\, \psi'(a)  \cr
\psi(b)+\rmi\,\psi'(b) } ,
\nonumber \\
& & \rho_2: \mathrm{D}(H^\dag)\to\mathbb{C}^2:\psi\mapsto \pmatrix{\psi(a)+\rmi\,\psi'(a)\cr \psi(b)-\rmi\,\psi'(b)}. 
\end{eqnarray}
These are well defined  since $\mathcal{H}^2(I)\subset C^1(I)$, and the following identity holds
\begin{equation}
\langle\rho_1(\psi)|\rho_1(\varphi)\rangle_{\mathbb{C}^2}-\langle\rho_2(\psi)|\rho_2(\varphi)\rangle_{\mathbb{C}^2}= 2\rmi\,\Gamma_{H^\dag}(\psi,\varphi),
\label{eq:bt}
\end{equation}
where $\Gamma_{H^\dag}(\psi,\varphi)= \langle H^\dag \psi| \varphi\rangle - \langle \psi| H^\dag\varphi\rangle$ is the boundary form defined in~(\ref{eq:boundf}), which measures the ``lack of self-adjointness'' of the operator $H$. Here $\langle\xi|\eta\rangle_{\mathbb{C}^2} = \overline{\xi}_1 \eta_1 + \bar{\xi}_2\eta_2$ is the canonical scalar product of $\xi,\eta\in\mathbb{C}^2$, while $\langle\psi|\varphi\rangle= \int_I \bar{\psi} (x) \varphi(x) \rmd x$ denotes the scalar product of $\psi,\varphi\in L^2(I)$. 

Given these maps, we have by~(\ref{eq:bt}) that $(\mathbb{C}^2,\rho_1,\rho_2)$ is a boundary triple (see~\ref{appendix}) for the Hamiltonian~(\ref{eq:ham}), and all  self-adjoint extensions of $H$ are given by~(\ref{eq:exts}), which reads
\begin{eqnarray}
\label{eq:bc}
& & \mathrm{D}(H_\mathcal{U})= \{\psi\in\mathcal{H}^2(I) :\; \left(\mathbb{I}-\mathcal{U}\right)
\pmatrix{ \psi(a)  \cr
\psi(b) } = \rmi \left(\mathbb{I}+\mathcal{U}\right)\pmatrix{-\psi'(a)  \cr
\psi'(b) }\} ,
\nonumber \\
& & H_\mathcal{U}\psi=-\frac{1}{2m}\psi''. 
\end{eqnarray}
where $\mathcal{U}$ is a unitary 2 $\times$ 2 matrix. 

This is the result obtained in~\cite{Asorey1,Asorey3}, which expresses all possible self-adjoint extensions of the Laplacian in terms of unitaries on the Hilbert space of boundary data $\mathbb{C}^2$.
The choice of particular unitary matrices gives rise to some well-known boundary conditions, for example:
\begin{eqnarray}
&  \mathcal{U}=-\mathbb{I}, &\quad  \psi(a)=0=\psi(b),  \qquad\qquad\quad\quad\;\;\; \mbox{Dirichlet};\\
& \mathcal{U}= \mathbb{I}, & \quad \psi'(a)=0=\psi'(b),  \qquad\qquad\quad\quad\;\,\mbox{Neumann};\\
&\mathcal{U}=\sigma_1, &  \quad \psi(a)=\psi(b)\, ,\,\psi'(a)=\psi'(b),   \quad\quad\;\; \mbox{periodic}; \label{eq:periodic}\\ 
&\mathcal{U}=-\sigma_1, & \quad \psi(a)=-\psi(b)\,,\,\psi'(a)=-\psi'(b),  \quad \mbox{antiperiodic},  \label{eq:antiperiodic}
\end{eqnarray}
$\sigma_1$ being the first Pauli matrix.

\begin{figure}[tbp]
\centering
\includegraphics[width=0.6\columnwidth]{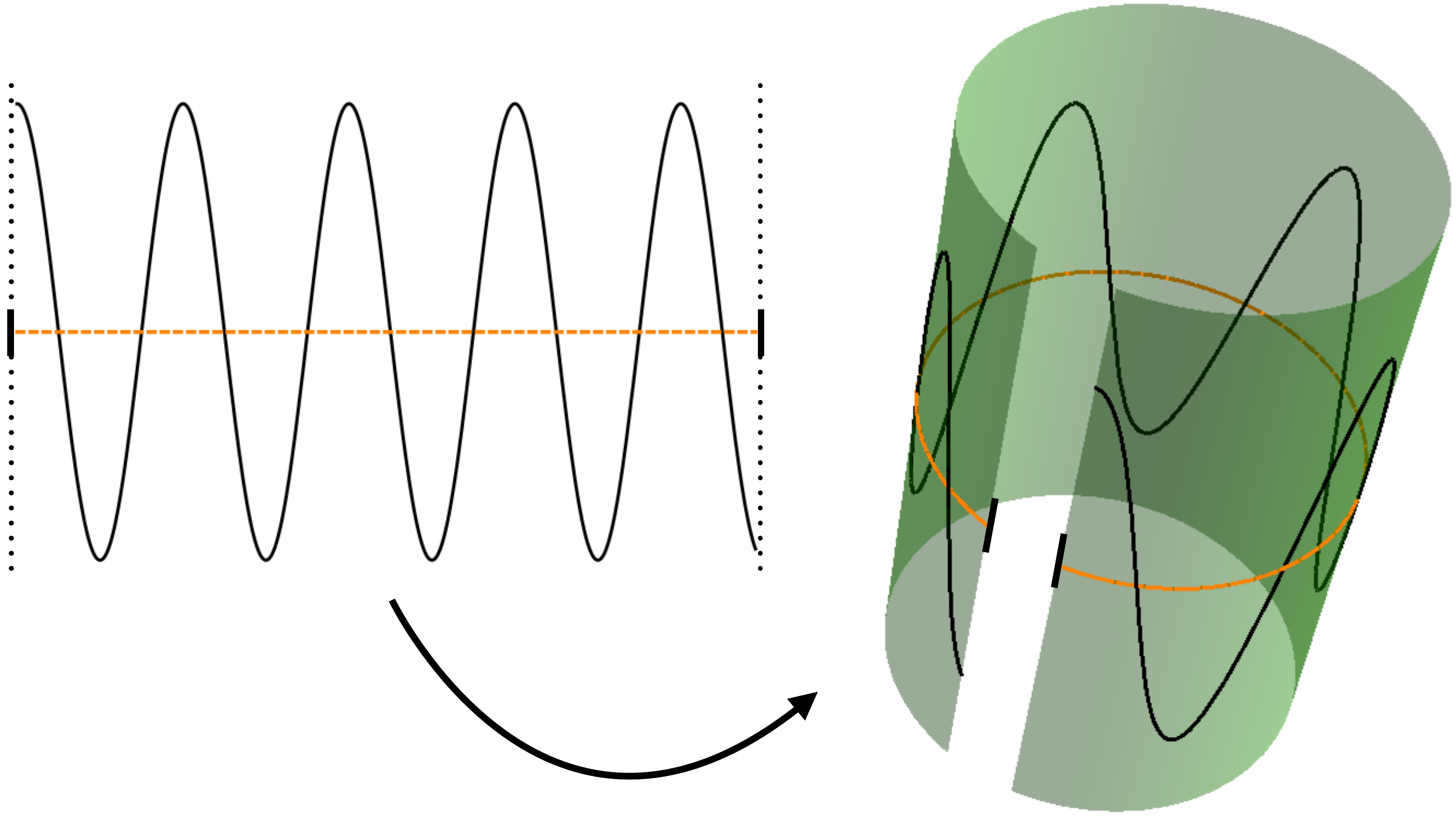}
\caption{After the bending, the functions defined over the interval transform as in figure.}
\label{fig:bendingfn}
\end{figure}

Now we would like to extract and parametrize a particular subset of boundary conditions 
which are invariant under dilations and will be useful in the following.
The set we are looking for is made up by all those boundary conditions which do not mix functions with derivatives at the boundary, that is of the form
\begin{equation}
\cases{\alpha\, \psi(a)+\beta\,\psi(b)=0,\\ \gamma\, \psi'(a)+\delta\, \psi'(b)=0,}
\end{equation}
where $\alpha,\beta,\gamma,\delta\in\mathbb{C}$.
It is easy to show that the conditions that have to be satisfied by the former four parameters  in order to represent a  self-adjoint extension of the Hamiltonian on the interval $I=[a,b]$ are
\begin{equation}
\beta \bar{\delta} = \alpha \bar{\gamma}.
\end{equation}
If we set $\eta=-\beta/\alpha$, the desired boundary conditions read
\begin{equation}
\label{eq:mybc}
\cases{\psi(a)=\eta	\, \psi(b),\\ 
\overline{\eta}\psi'(a)=\psi'(b),}
\end{equation}
and the unitary matrix  in~(\ref{eq:bc}) associated to this self-adjoint extension is provided by
\begin{equation}
\label{eq:non}
\mathcal{U}= \left(\begin{array}{cc} \frac{|\eta|^2-1}{1+|\eta|^2} & \frac{2\eta}{1+|\eta|^2}  \\ \frac{2\overline{\eta}}{1+|\eta|^2} & \frac{1-|\eta|^2}{1+|\eta|^2}  \end{array}\right) .
\end{equation}
Some comments are in order. If $\eta=\pm1$ we obtain periodic~(\ref{eq:periodic}) and antiperiodic~(\ref{eq:antiperiodic}) boundary conditions, while for $\eta=0$ or $\eta=\infty$ mixed Dirichlet and Neumann conditions arise. However, pure Dirichlet or Neumann conditions cannot be reached by our parametrization. Thus, the family in~(\ref{eq:non}), which we denote by $\{\mathcal{U}(\eta)\}_{\eta \in \mathbb{C}_{\infty}}$, where $\mathbb{C}_{\infty}=\mathbb{C}\cup \{\infty\}$, does not exhaust the whole set of dilation-invariant boundary conditions, which 
is instead provided by 
\begin{equation}
\label{eq:family}
\{\mathcal{U(\eta)_{\eta \in \mathbb{C}_{\infty}}}\,,\mathbb{I}\,,-\mathbb{I}\} .
\end{equation}
Moreover, it is worth noticing that the set $\{\mathcal{U}(\eta)\}$ 
does not form a subgroup of $\mathrm{U}(2)$. 

From a physical perspective the boundary conditions in~(\ref{eq:mybc}) are nonlocal, since they connect the value of the wave function at one end of the interval with its value at the other end. A physical realization of them require that the interval be bended into a ring with the two ends forming a tunneling junction through which the wave function can acquire a phase given by~(\ref{eq:non}). See Fig.~\ref{fig:bendingfn}. This  can be experimentally implemented by means e.g.\ of superconducting quantum interference devices, where the properties of the Josephson junction are suitably chosen to give the required phase~\cite{Asorey2}.

\section{Moving and fixed walls} \label{sec:movfix}
We start by generalizing the problem of a particle of mass $m$ in a one dimensional box with  moving walls subject to Dirichlet boundary conditions (extensively discussed in \cite{Sara,bangalectures})
to a larger class of boundary conditions, which we picked out in~(\ref{eq:family}).
For convenience we parametrize the one dimensional box by
\begin{equation}
I_{l,c}=[c-l/2, c+l/2],
\end{equation}
so that $c\in \mathbb{R}$ is the center of the interval, and $l>0$ is its length,
and consider the  Hamiltonian (kinetic energy)
\begin{eqnarray}
\fl\qquad H\psi&=-\frac{1}{2m}\psi'', \qquad \psi\in  \mathrm{D}_{l,c},
\nonumber\\
\fl\qquad \mathrm{D}_{l,c} &=\biggr\{\psi \in \mathrm{\mathcal{H}^2(}I_{l,c}\mathrm{)}:\; \psi\biggr(c-\frac{l}{2}\biggl)=\eta \psi\biggr(c+\frac{l}{2}\biggl)\,,\,\bar\eta\psi'\biggr(c-\frac{l}{2}\biggl)=\psi'\biggr(c+\frac{l}{2}\biggl)\biggl\},
\label{eq:myham}
\end{eqnarray}
where  $\eta$ is a fixed complex number representing particular boundary conditions~(\ref{eq:mybc}), and $\mathcal{H}^2(I_{l,c}\mathrm{)}$ is the Sobolev space of square integrable functions on {$I_{l,c}$,} whose first and second derivatives are square integrable functions.

Some comments are necessary. In the previous section we proved that the above boundary conditions yield a good self-adjoint extension of the Hamiltonian on an interval. As already remarked these do not mix the values of the functions at the border with their derivatives. In what follows we will see that these are the only ones which are invariant under dilations, a crucial property for what we are going to investigate.

Next we take into account the dynamics of this problem by taking smooth paths in the parameter space $(l,c)\in \mathbb{R}_+\times\mathbb{R}$: $t\mapsto c(t)$ and $t\mapsto l(t)$. 
Clearly we are translating the box by $c(t)$ and contracting/dilating it by $l(t)$.
As underlined in~\cite{Sara} determining the quantum dynamics of this system is not an easy problem to tackle with, since we have Hilbert spaces, $L^2(I_{l(t),c(t)})$, varying with time and we need to compare vectors in different spaces.
The standard approach is to embed the time-dependent spaces into a larger one, namely $L^2(\mathbb{R})$, extend the two-parameter family of Hamiltonians~(\ref{eq:myham}) to this space and try to unitarily map the problem we started with into another one, with a family of time-dependent Hamiltonians on a fixed common domain. 

With this end in view we embed $L^2(I_{l,c})$ into $L^2(\mathbb{R})$ in the following way
\begin{equation}
L^2(\mathbb{R})=L^2(I_{l,c})\oplus L^2( I_{l,c}^\mathrm{c}),
\end{equation}
where $I^\mathrm{c} = \mathbb{R}\setminus I$ is the complement of the set $I$,
so that we can consider the extension of the Hamiltonians defined in~(\ref{eq:myham}) as
\begin{equation}
\label{eq:embham}
H(l,c)=\frac{p^2}{2m}{\oplus}_{l,c} \bm{0},
\end{equation}
where the embedding and the direct sum obviously depend on $l$ and $c$.
Following~$\cite{Sara}$ we recall how to reduce this moving walls problem into a fixed domain one. 
The composition of a translation $x\rightarrow x-c$ and of a subsequent dilation $x\rightarrow x/l$ maps the interval $I_{l,c}$ onto 
\begin{equation}
I=I_{1,0}=\left[-\frac{1}{2},\frac{1}{2}\right],
\end{equation} 
which does not depend on $c$ and $l$. Next we need to define a unitary action of both groups on $L^2(\mathbb{R})$.
A possible choice is
\begin{equation}
\label{eq:2.1}
\fl \qquad (V(c)\psi)(x)=\psi(x-c), \qquad (W(s)\psi)(x)=\rme^{-s/2}\psi(\rme^{-s}x),\quad \forall \psi \in L^2(\mathbb{R}),
\end{equation}
and both $c\in\mathbb{R}\rightarrow V(c)$ and $s =\ln l\in\mathbb{R}\rightarrow W(s)$ form  one-parameter (strongly continuous) unitary groups. 
The factor $\exp(-s/2)$ is consistent with the physical expectation that $\psi$ transforms as the square root of a density under dilation.

In order to make the expression $\ln l$ meaningful, from now on we are going to identify $l$ with a pure number given by the ratio of the actual length of the box and a unitary length.
The infinitesimal generator of the group of translations is the momentum operator
\begin{equation}
\label{eq:momentum}
p: \mathrm{D}(p) = \mathcal{H}^1 (\mathbb{R})\rightarrow L^2(\mathbb{R}) , \qquad p\psi=-\rmi\,\psi',
\end{equation}
so that spatial translations are implemented by the unitary group
\begin{equation}
V(c)=\mathrm{exp}\left(-\rmi\, c\, p\right), \qquad \forall\,c\in\mathbb{R}.
\label{eq:Vtransl}
\end{equation}
Similarly, the generator of the dilation unitary group is given by the virial operator over its maximal domain:
\begin{eqnarray}
\fl \qquad x\circ p:=\overline{xp-\frac{\rmi}{2}}=\overline{\frac{1}{2}(xp+px)},
\qquad
\mathrm{D}(x\circ p)=\{\psi \in L^2(\mathbb{R})\,|\,x\psi' \in L^2(\mathbb{R})\},
\end{eqnarray}
where $\overline{A}$ denotes the closure of the operator $A$. Dilations on $L^2(\mathbb{R})$ are, thus, implemented by
\begin{equation}
W(s)=\mathrm{exp}\left(-\rmi\,s\, x \circ p\right), \qquad \forall\,s\in\mathbb{R}.
\end{equation}
Next we define the two-parameter family of unitary operators on $L^2(\mathbb{R})$, which are going to fix our time-dependent problem
\begin{equation}
\label{eq:diltr}
U(l,c): L^2(\mathbb{R})\rightarrow L^2(\mathbb{R}), \qquad U(l,c)=W^\dag(\ln l) V^\dag(c).
\end{equation}
By this unitary isomorphism we are mapping $H(l,c)$ into
\begin{equation}
\label{eq:mynewham}
H(l)= U(l,c)H(l,c)U^\dag(l,c)=\frac{p^2}{2 m l^2} \oplus \bm{0},
\end{equation}
where we have used the identity
\begin{equation}
\label{eq:core}
W^\dag(\ln l)pW(\ln l)= \frac{p}{l}. 
\end{equation}
The operators in~(\ref{eq:mynewham}) act on the time-independent  domain
\begin{equation}
\mathrm{D}(H(l))=\mathrm{D}\oplus L^2(I^c),
\end{equation}
where $\mathrm{D}=U(l,c)\mathrm{D}_{l,c}$ is given by
\begin{equation}
\label{eq:dom}
\mathrm{D}=\biggr\{\psi \in \mathrm{H}^2(I)\,:\,\psi\biggr(\!\!-\frac{1}{2}\biggl)=\eta \psi\biggr(\frac{1}{2}\biggl)\,,\,\bar\eta\psi'\biggr(\!\!-\frac{1}{2}\biggl)=\psi'\biggr(\frac{1}{2}\biggl)\biggl\}.
\end{equation}
We have thus achieved our goal, that is mapping the initial family of Hamiltonians with time-dependent domains into a family with a common fixed domain of self-adjointness. This has been possible thanks to the unitary operator~(\ref{eq:diltr}) and, most importantly, to the choice of dilation-invariant boundary conditions~(\ref{eq:myham}) as discussed in the previous section. We have taken into account those boundary conditions~(\ref{eq:mybc}) which do not mix derivatives and functions at the boundary: these are the only ones which leave the transformed domain $\mathrm{D}=U(l,c)\mathrm{D}_{l,c}$ in~(\ref{eq:dom}) time-independent.

\section{The Berry phase factor} \label{sec:Berry}
The main objective of this section will be to exhibit a non-trivial geometric phase associated to a cyclic adiabatic evolution of the physical system described in~(\ref{eq:myham}). 
Let $\mathcal{C}$ be a closed path in the parameter space $(l,c)\in\mathbb{R}_+\times\mathbb{R}$. Let the $n$-th energy level be non degenerate; then, in the adiabatic approximation, the Berry phase associated to the cyclical adiabatic evolution is given by
\begin{equation}
\label{eq:Berry}
\Phi_n=\oint_{\mathcal{C}} \mathcal{A}^{(n)}=\rmi\oint_{\mathcal{C}}\langle\psi_n | \rmd\psi_n\rangle , 
\end{equation}
where $\psi_n$ is the eigenfunction associated to the $n$-th eigenvalue, $\rmd$ is the external differential defined over the parameter manifold $\mathbb{R}_+\times\mathbb{R}$, and 
\begin{equation}
\label{eq:integ}
\langle\psi_n|\rmd\psi_n\rangle= \int_{\mathbb{R}}\overline{\psi_n(x)}(\rmd\psi_n)(x) \rmd x.
\end{equation}
In our  case $\rmd\psi_n$ reads
\begin{equation}
\label{eq:diff}
(\rmd\psi_n)(x)=\biggl(\frac{\partial}{\partial l}\psi_n\biggr)(x) \rmd l+\biggl(\frac{\partial}{\partial c}\psi_n\biggr)(x)\rmd c.
\end{equation}

A technical difficulty arises from equations~(\ref{eq:Berry})-(\ref{eq:diff}). In this section we are going to show that, for fixed $\eta$, the eigenfunctions $\{\psi_n\}_{n\in\mathbb{N}}$ determine an orthonormal basis in $L^2(\mathbb{R})$. However, in general the derivatives in~(\ref{eq:diff}) do not belong to $L^2(\mathbb{R})$  so that the integral in~(\ref{eq:integ}) is ill-posed and needs a prescription of calculation. No doubt that the ill-posedness of~(\ref{eq:Berry}) is due to the presence of a boundary in our system.
  
First we need to determine the spectral decomposition of the Hamiltonian we started with in~(\ref{eq:myham}) or equivalently in~(\ref{eq:embham}). Of course this would be a difficult problem to handle, but thanks to the unitary operator in~(\ref{eq:diltr}) we can move on to the Hamiltonians with fixed domain, compute the spectral decomposition and then make our way unitarily back to the  problem with time-dependent domain.
Therefore,  we need to solve the  eigenvalue problem
\begin{equation}
-\frac{1}{2m l^2}\phi''(x)=\lambda \phi(x),
\end{equation}
where $\phi\in \mathrm{D}$ in~(\ref{eq:dom}) and $\lambda \in \mathbb{R}$.
The spectral decomposition will heavily rely on the choice of the parameter $\eta$, which, as already 
stressed, represents a particular choice of boundary condition. If $\eta \ne \pm 1$ the spectrum is non-degenerate, 
and the normalized eigenfunctions have the form
\begin{equation}
\phi_n(x)=\sin(k_n x) + \rme^{\rmi\,\alpha}\cos(k_n x), \qquad n\in\mathbb{Z},
\end{equation}
where
\begin{eqnarray}
\alpha=\mathrm{Arg}\biggl(\frac{1+\eta}{1-\eta}\biggr),
\qquad
k_n = 2n\pi + 2 \arctan\biggl|\frac{1-\eta}{1+\eta}\biggr|,
\qquad n\in\mathbb{Z},
\label{eq:eigenvalues}
\end{eqnarray}
so that the dispersion relation ($\lambda=k^2/2 m l^2$) reads
\begin{equation}
\lambda_n =\frac{2}{m l^2}\biggl(n\pi + \arctan\biggl|\frac{1-\eta}{1+\eta}\biggr| \biggr)^2,\qquad n\in\mathbb{Z}.
\end{equation}

Let $\varphi$ be an arbitrary function belonging to $L^2(I^{\mathrm{c}})$; $\varphi$ is clearly an eigenfunction of the \textbf{0} operator on $L^2(I^{\mathrm{c}})$ with  zero eigenvalue. Therefore we can extend $\phi_n$ to an eigenfunction of~(\ref{eq:embham}), $\psi_n=\phi_n\oplus\varphi$, which can be conveniently chosen to be a test function:
$\psi_n\in\mathcal{D}(\mathbb{R})$, the space of smooth functions with compact support. 
\begin{figure}[tbp]
\centering
\includegraphics[width=0.45\columnwidth]{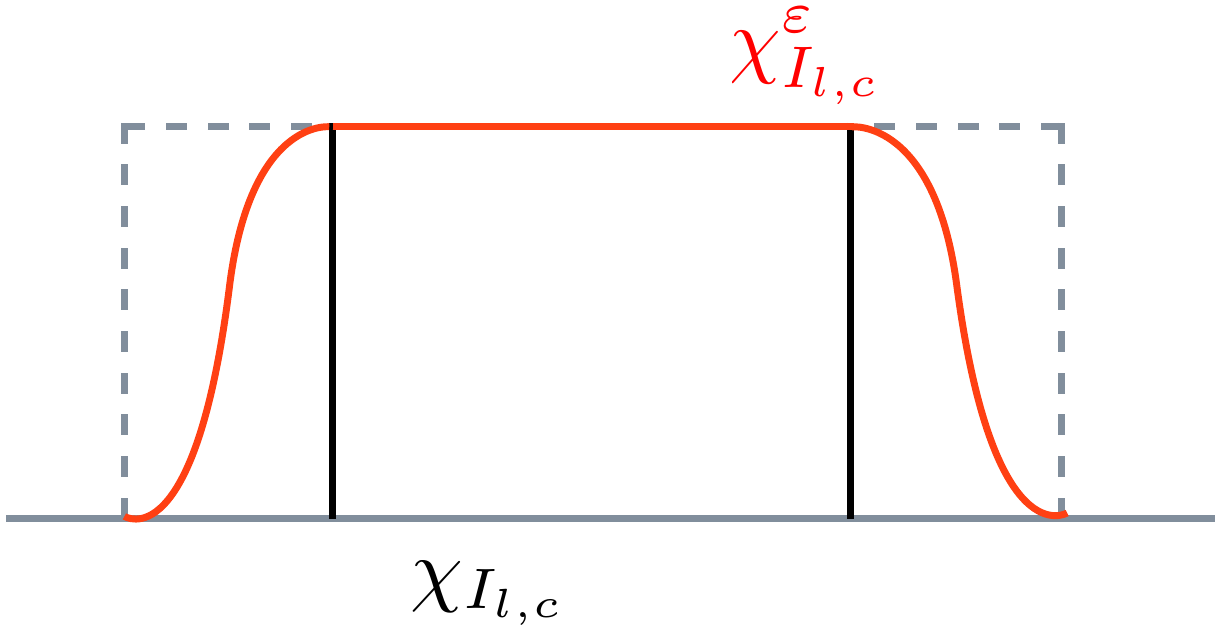}
\caption{Regularized characteristic function~(\ref{eq:regchar}).}
\label{molly2}
\end{figure}
For the sake of the reader, let us  exhibit an explicit construction of $\psi_n(x;l,c)$. 

Let $\tilde{\phi}_n(x;l,c)$ be a smooth extension of $\phi_n(x;l,c)\in\mathrm{D}_{l,c}\subset L^2(I_{l,c})$ to the whole real line. Roughly speaking our eigenfunction can be written as the restriction of this extension, namely $\tilde{\phi}_n(x;l,c)\chi_{I_{l,c}}(x)$, where $\chi_A(x)$ is the characteristic function of the set $A$ [$\chi_A(x)=1$ if $x\in A$, and $=0$ otherwise], showing why divergent contributions arise from the boundary when taking  derivatives. 
So the idea which underlies the following discussion is to regularize the contribution of the characteristic function $\chi_{I_{l,c}}$. 

\begin{figure}[tbp]
\centering
\includegraphics[width=0.55\columnwidth]{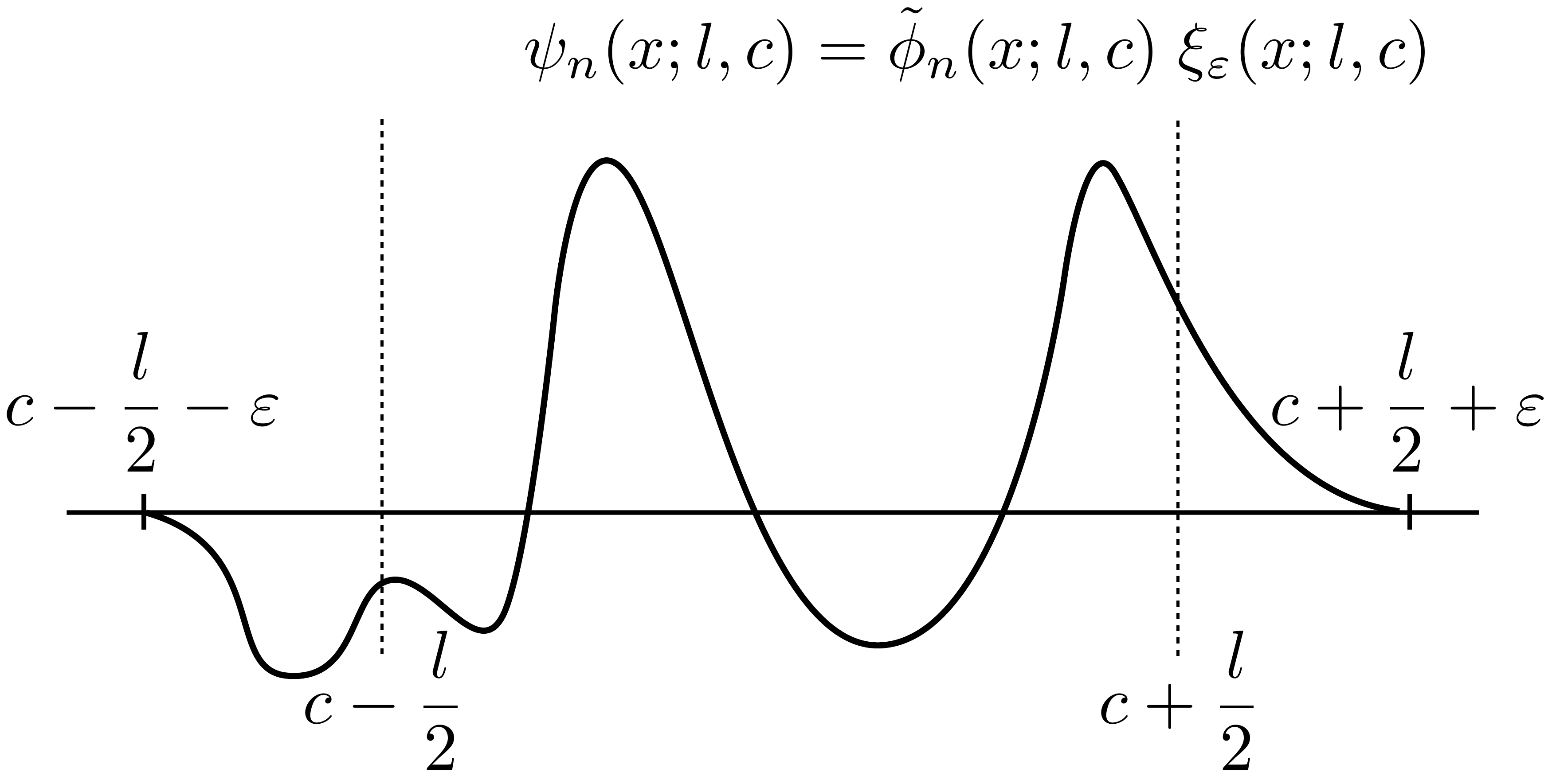}
\caption{The regularization procedure~(\ref{eq:emb}).}
\label{molly}
\end{figure}

Let $\rho(x)$ be a nonnegative monotone decreasing  function which belongs to $C^{\infty}([0,\infty))$, moreover we require that $\rho(0)=1$, $\rho(1)=0$ and $\rho^{(n)}(0)=0$ for $n\geq1$. 
We are going to paste two contracted  copies of the latter to $\chi_{I_{l,c}}$, such that the final result would be as in Figure~\ref{molly2}.
Given $\varepsilon>0$ we define the regularized characteristic function of $I_{l,c}$ as follows:
\begin{equation}
\chi^\varepsilon_{I_{l,c}}(x)=
\cases{1&for $x\in I_{l,c}$\\
\rho\left(\frac{|x-c|- l/2}{\varepsilon}\right)& for $x\not\in I_{l,c}$ },
\label{eq:regchar}
\end{equation}
which is a test function, $\chi^\varepsilon_{I_{l,c}}\in \mathcal{D}(\mathbb{R})$.
In light of the previous discussion we choose the following functions and show that they are eigenfunctions for the Hamiltonian~(\ref{eq:embham}):
\begin{equation}
\label{eq:emb}
\psi_n(x;l,c)=\tilde{\phi}_n(x;l,c)\; \xi_\varepsilon(x;l,c),\qquad \varepsilon > 0,
\end{equation}
where
\begin{equation}
\xi_\varepsilon(x;l,c)=\frac{1}{{\|\tilde{\phi}_n\; \chi^\varepsilon_{I_{l,c}}\|}}\chi^\varepsilon_{I_{l,c}}(x).
\label{eq:normchar}
\end{equation}
See Figure~\ref{molly}.

Even if $\tilde{\phi}_n\notin L^2(\mathbb{R})$,  this will not alter the desired regularity property and the integrability condition of~(\ref{eq:emb}). Clearly~(\ref{eq:emb}) is still an eigenfunction of~(\ref{eq:embham}) because $\psi_n|_{I_{l,c}}=\phi_n$ is an eigenfunction of the Hamiltonian  defined in~(\ref{eq:myham}) and $\psi_n|_{I_{l,c}^{\mathrm{c}}}$ is trivially an eigenfunction of the \textbf{0} operator with null eigenvalue.
Moreover, from the explicit expression in~(\ref{eq:normchar}) this eigenfunction is normalized.

In this renormalization scheme, which is needed for the definiteness of~(\ref{eq:diff}),  we are  first  embedding $\mathrm{D}_{l,c}\subset L^2(I_{l,c})$ into $L^2(\mathbb R)$ and then regularizing the boundary contribution through the introduction of the regularizer $\rho$. 

Now it is essential to observe that
\begin{equation}
\lim_{\varepsilon\to0} \tilde{\phi}_n(x;l,c)\,\xi_\varepsilon(x;l,c) =\tilde{\phi}_n(x;l,c)\,\chi_{I_{l,c}}(x)=\phi_n(x;l,c) \oplus 0,
\end{equation}
that is the eigenfunction of a particle confined in $I_{l,c}$. Here, the convergence of the limit is pointwise and, by dominated convergence, in $L^2(\mathbb{R})$.

We are now in the right position to compute~(\ref{eq:integ}) for $\varepsilon>0$, which is well posed, and then take the limit $\varepsilon\to 0$.
We start by considering separately both the terms in
\begin{equation}
\langle\psi_n|\rmd\psi_n\rangle=\biggl(\int_{\mathbb{R}}\overline{\psi_n(x)}\frac{\partial}{\partial l}\psi_n(x)\rmd x\biggr)\rmd l+\biggl(\int_{\mathbb{R}}\overline{\psi_n(x)}\frac{\partial}{\partial c}\psi_n(x)\rmd x\biggr)\rmd c,
\end{equation}
which, after an integration by parts, become
\begin{eqnarray}
\label{eq:coefficients}
& &\int_{\mathbb{R}}\overline{\psi}_n\frac{\partial}{\partial l}\psi_n \rmd x=\frac{1}{2}\int_{\mathbb{R}}\frac{\partial}{\partial l}|\psi_n|^2\rmd x+\rmi\,\mathrm{Im}\int_{\mathbb{R}}\biggl(\overline{\phi}_n\frac{\partial}{\partial l}\phi_n\biggr) \xi_\varepsilon^2 \rmd x,\\
& & \int_{\mathbb{R}}\overline{\psi}_n\frac{\partial}{\partial c}\psi_n \rmd x=\frac{1}{2}\int_{\mathbb{R}}\frac{\partial}{\partial c}|\psi_n|^2\rmd x+\rmi\,\mathrm{Im}\int_{\mathbb{R}}\biggl(\overline{\phi}_n \frac{\partial}{\partial c}\phi_n\biggr) \xi_\varepsilon^2 \rmd x.
\end{eqnarray}
By plugging the explicit expressions of the eigenfunctions we find, by dominated convergence, that for $\varepsilon\to 0$
\begin{eqnarray}
\fl \quad & & \mathrm{Im}\int_{\mathbb{R}}\biggl(\overline{\phi}_n \frac{\partial}{\partial l}\phi_n\biggr) \xi_\varepsilon^2 \rmd x=\frac{k_n}{l^3}\sin\alpha\int_{\mathbb{R}}(x-c)\xi_\varepsilon^2(x) \rmd x \to 
\frac{k_n}{l^3}\sin\alpha \int_{I_{l,c}}(x-c)
\rmd x=
0,\\
\fl\quad & & \mathrm{Im}\int_{\mathbb{R}}\biggl(\overline{\phi}_n\frac{\partial}{\partial c}\phi_n\biggr) \xi_\varepsilon^2 \rmd x=\frac{k_n}{l^2}\sin\alpha\int_{\mathbb{R}}\xi_\varepsilon^2(x) \rmd x
\to 
\frac{k_n}{l^2}\sin\alpha \int_{\mathbb{R}}\chi_{I_{l,c}}(x)\rmd x=
\frac{k_n}{l}\sin\alpha.
\end{eqnarray}
Moreover, since $\psi_n$ has inherited from $U^\dag(l,c)$ the right regularity properties, 
for any $\varepsilon>0$, one gets
\begin{equation}
\fl \qquad \int_{\mathbb{R}}\frac{\partial}{\partial l}|\psi_n|^2\rmd x=\frac{\partial}{\partial l}\int_{\mathbb{R}}|\psi_n|^2\rmd x=0, \qquad
\int_{\mathbb{R}}\frac{\partial}{\partial c}|\psi_n|^2\rmd x=\frac{\partial}{\partial c}\int_{\mathbb{R}}|\psi_n|^2\rmd x=0.
\end{equation}

\begin{figure}[tbp]\centering
\includegraphics[width=0.4\textwidth]{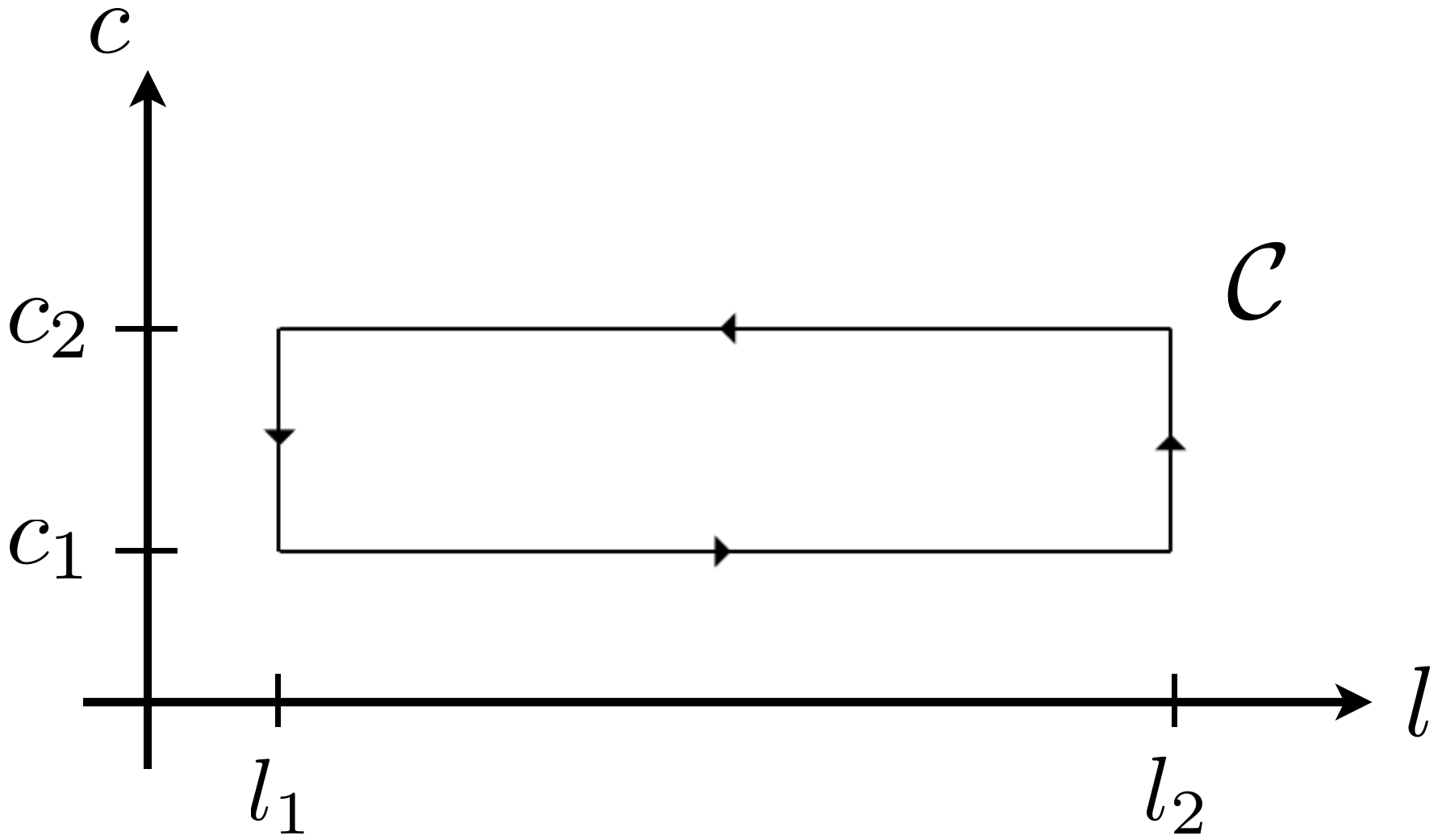}
\caption{The adiabatic path $\mathcal{C}$.}
\label{fig:path}
\end{figure}

Summing up, we finally get the expression of the Berry one-form:
\begin{equation}
\label{eq:form}
\langle\psi_n | d\psi_n\rangle=\rmi\biggl(\frac{k_n}{l}\sin\alpha\biggr)\rmd c,
\end{equation}
which is manifestly not closed yielding a nontrivial Abelian phase. Notice that the one-form derived in~(\ref{eq:form}) is purely imaginary, consistently with the general theory of Berry phases~\cite{Bohm}. Moreover it heavily depends on the energy level through $k_n$ in~(\ref{eq:eigenvalues}) and on the boundary conditions through $\mathrm{sin}\,\alpha$.

\begin{figure}[tbp]\centering
\includegraphics[width=0.5\textwidth]{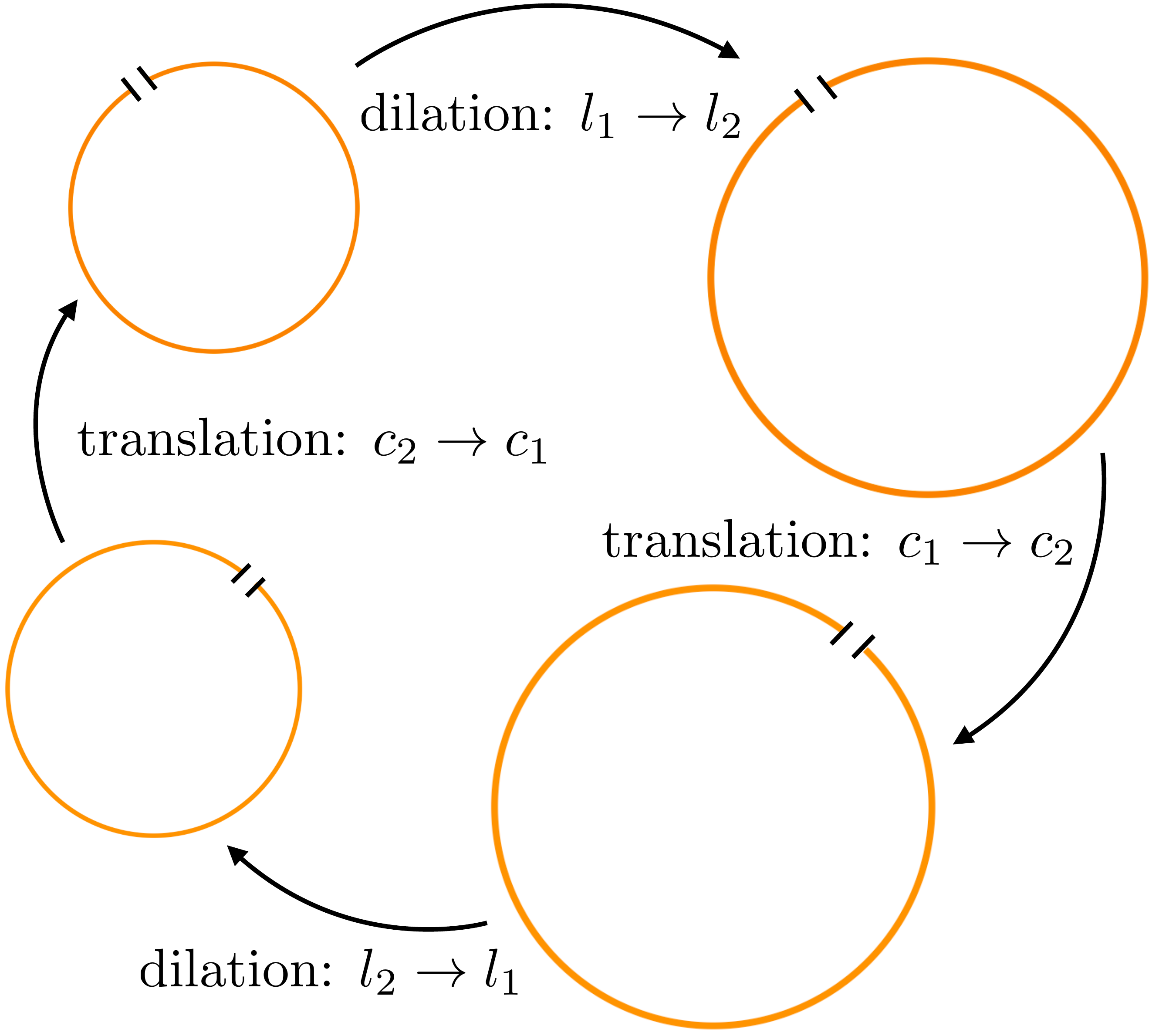}
\caption{Cyclic evolution according to the path drawn in Figure~\ref{fig:path}.\label{fig:circles}}
\end{figure}

As a simple example, we choose a rectangular path $\mathcal{C}$ in the $(l,c)$ half-plane, as shown in Figure~\ref{fig:path}, and compute
\begin{equation}
\Phi_n=\oint_{\mathcal{C}} \mathcal{A}^{(n)}=\rmi \oint_{\mathcal{C}}\langle\psi_n | \rmd \psi_n\rangle, 
\end{equation}
whose only non-trivial contributions are given by the vertical components of the circuit.
The final result is
\begin{equation}
\Phi_n=\oint_{\mathcal{C}} \mathcal{A}^{(n)}=k_n\biggl(\frac{1}{l_1}-\frac{1}{l_2}\biggr)(c_2-c_1) \sin\alpha,
\end{equation}
which, as expected, depends on the particular path chosen. In the spirit of the physical implementation of our system in terms of a ring with a junction (see section~\ref{BoundCond}), our cyclic adiabatic evolution could be illustrated as in Figure~\ref{fig:circles}.

Another interesting aspect provided by this problem is linked to a nontrivial Berry curvature:
\begin{equation}
\label{eq:curv}
\mathcal{F}^{(n)}=\rmd \mathcal{A}^{(n)}=\frac{k_n}{l^2}\sin\alpha\; \rmd l \wedge \rmd c.
\end{equation}

The above formula brings to mind the curvature of a hyperbolic Riemannian manifold. Indeed, consider the \textit{Poincar\'e half-plane}, which by definition is the upper-half plane together with the Poincar\'e metric:
\begin{equation}
\rmd s^2=\frac{\rmd x^2+\rmd y^2}{y^2}.
\end{equation}
The half-plane is a model of hyperbolic geometry and if we consider the area form on it we have 
\begin{equation}
A=\frac{\rmd x\wedge \rmd y}{y^2},
\end{equation}
which has the same structure as the Berry curvature~(\ref{eq:curv}) of our quantum mechanical model.

We remark that the relevant group in hyperbolic geometry is $\mathrm{PSL}(2,\mathbb{R})$ the group of real M\"obius transformations. 
The Lie algebra of this group is the space of real $2\times2$ traceless matrices which are spanned by $\sigma_1,\sigma_3$ and $i\sigma_2$,
where the $\sigma$s are the usual Pauli matrices. The two generators $\sigma_3$ and $\sigma_+=\sigma_1+i\sigma_2$ form a closed subalgebra. 
The structure of this Lie subalgebra is exactly the same as ours: the commutator of the virial operator and the momentum operator is the momentum operator, namely,
\begin{equation}
[x\circ p, p]= i p.
\end{equation}

\section{The regularization procedure: an equivalent perspective} \label{sec:regularization}
One may object to the regularization scheme introduced in the previous section for its artificiality. In fact, in order to have a well-posed problem we embedded our original problem into a larger space, $L^2({\mathbb{R}})$, and needed to make sense of the differential in~(\ref{eq:diff}). In this section we are going to understand better what may be the problem in the definition of the derivative with respect to our parameters, and, moreover, we are going to show an alternative, intrinsic, approach to renormalization which does not make use of any embedding.
Let us consider the following map:
\begin{equation}
\label{eq:unit}
(l,c)\in\mathbb{R_{+}}\times\mathbb{R}\mapsto \zeta(l,c)=U^\dag(l,c)\zeta=V(c)\,W(\ln l) \zeta\in L^2(\mathbb{R}),
\end{equation}
where $\zeta\in L^2(\mathbb{R})$ is a suitable  unit vector independent of $(l,c)$, and $U(l,c)$ is defined in~(\ref{eq:diltr}). 
We would like to understand better the following differential:
\begin{equation}
(\rmd \zeta)(x)=\biggl(\frac{\partial}{\partial l}\zeta\biggr)(x) \rmd l+\biggl(\frac{\partial}{\partial c}\zeta\biggr)(x)\rmd c.
\end{equation}
Fix $l>0$ and consider the restriction of~(\ref{eq:unit}) to its second argument
\begin{equation}
c\in\mathbb{R}\mapsto U^\dag(l,c)\zeta=V(c)\, W(\ln l)\zeta.
\end{equation}
$\{V(c)\}_{c\in\mathbb{R}}$ in~(\ref{eq:Vtransl}) form a one-parameter group, whose generator is the momentum $p$ defined in~(\ref{eq:momentum})
Thus,
\begin{equation}
\frac{\partial}{\partial c}\zeta (l,c)=\left(\frac{\rmd }{\rmd c} V(c)\right) \left(W(\ln l)\zeta\right)= -\rmi\,p V(c) \left(W(\ln l)\zeta\right),
\end{equation}
which is well posed if and only if $W(\ln l)\zeta \in \mathcal{H}^1(\mathbb{R})$. For this reason we can interpret $\partial \zeta (l,c)/ \partial c$ as the distributional derivative of $\zeta (l,c)$ which is forced to belong to $L^2(\mathbb{R})$.
In our case the extension of the eigenfunction to the real line is smooth, $\tilde\phi_n(l,c) \in C^{\infty}(\mathbb{R})$, so that the derivatives can be computed classically. Clearly $\partial \tilde\phi_n (l,c)/ \partial c$ is only locally summable over the real line. Since the restriction of smooth functions to open subsets is still smooth and since from a physical perspective we can have information only on what happens on the inside of the one dimensional box, $I_{l,c}$, we give the following prescription: 
\begin{equation}
 \frac{\partial}{\partial c}   \phi_n (l,c) := \frac{\partial}{\partial c}   \tilde\phi_n (l,c) \Big|_{\mathring{I}_{l,c}}
\end{equation}
being an element of $C^{\infty}(\mathring{I}_{l,c})$ and locally summable. An analogous prescription works for the derivative with respect to \textit{l}.
Let us return to our problem settled in $\mathcal{H}= L^2(I_{l,c})$. This time the one-form  is given by
\begin{equation}
\label{eq:distr}
\fl \quad \langle\phi_n | \rmd \phi_n\rangle=\biggl(\int_{I_{l,c}}\overline{\phi_n(x)}\left(\frac{\partial}{\partial l}\phi_n\right)(x)\rmd x\biggr)\rmd l+\biggl(\int_{I_{l,c}}\overline{\phi_n(x)}\left(\frac{\partial}{\partial c}\phi_n\right)(x)\rmd x\biggr)\rmd c,
\end{equation}
where the derivatives in~(\ref{eq:distr}) are to be considered in the sense stated above, that is as locally integrable functions in $\mathring{I}_{l,c}$. 
Once more,
\begin{equation}
\label{abph:19}
\int_{I_{l,c}}\overline{\phi_n(x)}\frac{\partial}{\partial l}\phi_n(x) \rmd x=\int_{I_{l,c}}\frac{\partial}{\partial l}|\phi_n(x)|^2\rmd x- \int_{I_{l,c}}\phi_n(x)\frac{\partial}{\partial l}\overline{\phi_n(x)}\rmd x.
\end{equation}
Due to normalization the first factor in the second member vanishes so that
\begin{equation}
\label{abph:20}
\mathrm{Re}\left(\int_{I_{l,c}}\overline{\phi_n(x)}\frac{\partial}{\partial l}\phi_n(x) \rmd x\right)=0,
\end{equation}
while  as  before we get
\begin{equation}
\int_{I_{l,c}}\overline{\phi_n(x)}\frac{\partial}{\partial l}\phi_n(x) \rmd x= \rmi\,\mathrm{Im}\int_{I_{l,c}}\overline{\phi_n(x)}\frac{\partial}{\partial l}\phi_n(x) \rmd x,
\end{equation}
and an analogous expression for the partial derivative with respect to $c$ holds.

With this in mind we are able to get the same result~(\ref{eq:form}) as before, by
reaching the boundary from the ``inside'', rather than from the  ``outside", so that our new prescription, though equivalent to the one  discussed above, may appear more natural. This is coherent from a physical perspective since we can have information only on what happens on the inside of the one dimensional box $I_{l,c}$.

\section{The degenerate case} \label{sec:degen}
For completeness, we are going to investigate the exceptional cases $\eta=\pm1$, which, as mentioned before, correspond to degenerate spectra.
For $\eta=1$ we have that for any $n\geq1$ the two eigenvalues $\lambda_{n}$ and $\lambda_{-n}$
in~(\ref{eq:eigenvalues}) coalesce, 
and an orthonormal basis in the $n$-th eigenspace  is given by
\begin{equation}
\label{eq:basis}
\phi^I_n(x) 
=\sqrt{2}\,\cos(2{\pi}n\,x) ,
\qquad
\phi^{II}_n(x) 
=\sqrt{2}\,\sin(2{\pi}n\,x), \qquad n\geq 1.
\end{equation}
For $\eta=-1$, we have instead that $\lambda_n=\lambda_{-n-1}$,
and a possible choice of an orthonormal basis is
\begin{equation}
\fl\qquad \phi^I_n(x) 
=\sqrt{2}\,\cos((2n+1){\pi}\,x) ,
\qquad \phi^{II}_n(x) 
=\sqrt{2}\,\sin((2n+1){\pi}\,x), \qquad n\in\mathbb{N}.
\end{equation}

From the general theory of geometric phases~\cite{wilczekzee} it is well known that a degenerate spectral decomposition gives rise to a one-form connection in terms of a Hermitian matrix and from a geometrical perspective this corresponds to a connection on a principal bundle, whose typical fiber is identified with a non-Abelian group. 

Let us consider the case $\eta=1$, which physically corresponds to periodic boundary conditions. We need to compute the following matrix one-form:
\begin{equation}
\label{eq:nonm}
\mathcal{A}^{(n)}=\rmi\left(\begin{array}{cc} \langle\phi_n^I | \rmd \phi_n^I\rangle & \langle\phi_n^I | \rmd \phi_n^{II}\rangle  \\ 
\langle\phi_n^{II} | \rmd \phi_n^I\rangle & \langle\phi_n^{II} | \rmd \phi_n^{II}\rangle  \end{array}\right) ,
\end{equation}
where the coefficients of the differentials are to be considered in the distributional sense.
The former equation yields the following result:
\begin{equation}
\label{eq:onef}
\mathcal{A}^{(n)}=\mathcal{A}_l^{(n)}\rmd l+\mathcal{A}_a^{(n)}\rmd c=\frac{k_n}{l}\sigma_2\rmd c.
\end{equation}
where $\sigma_2$ is the second Pauli matrix.
For a non-Abelian principal fiber bundle, the curvature two-form, according to the Cartan structure equation, is provided by
\begin{equation}
\mathcal{F}^{(n)}=\rmd \mathcal{A}^{(n)}+\frac{1}{2}[\mathcal{A}^{(n)},\mathcal{A}^{(n)}].
\end{equation}
Plugging in the explicit expression of the above one-form~(\ref{eq:onef}) we find that
\begin{equation}
\mathcal{F}^{(n)}=\rmd \mathcal{A}^{(n)}.
\end{equation}
The latter equation shows explicitly that, although every fiber is two dimensional, the overall bundle is trivial.
The one-form connection in~(\ref{eq:onef}) can be \textit{globally} diagonalized making use of the basis of plane waves. Indeed, if we had started from a ``rotated" basis, instead of~(\ref{eq:basis}):
\begin{equation}
\phi_n^I(x)\pm\,\rmi\,\phi_n^{II}(x)\propto \rme^{\pm\rmi\,k_n x},
\end{equation}
due to Euler's identity, and computed~(\ref{eq:onef}) in this new basis, we would have obtained a diagonal matrix.
In the most general case, instead, one is able to determine only a \textit{local} basis where the above one-form~(\ref{eq:onef}) is diagonal. On the other hand, in our case the bundle can be globally trivialized.

\section{Conclusions}  \label{sec:conclusion}
We have considered the problem of a particle in a box with moving walls with a class of boundary conditions. Unlike the example studied by Berry and Wilkinson (two dimensional region with 
Dirichlet boundary conditions), our box is one dimensional and we impose more general boundary 
conditions. We consider situations in which the location and the size of the box 
are slowly varied. Our problem is complicated by the fact that different points in the parameter space correspond to different Hilbert spaces. In order to deal with this we need to
invoke a larger Hilbert space and exercise care while varying our two parameters.
Within this two parameter space we conclude that there is a non-trivial geometric phase. The functional form of this phase two-form is suggestive of the area two-form 
in hyperbolic geometry. 

Our boundary conditions in general violate time reversal symmetry, i.e, the complex conjugate of a wave function which satisfies the boundary condition described by $\eta$ 
may not satisfy the same boundary condition. In fact, the only boundary conditions that respect time reversal are those where $\eta$ is real. In this case,
we would expect the geometric phase to reduce to the topological phase (which only takes values $\pm 1$). Thus the two form describing the phase must vanish.
In fact when $\eta$ is real (but not equal to $\pm 1$, which is a degenerate case), $\alpha$ in~(\ref{eq:eigenvalues}) is zero or $\pi$ and 
the corresponding geometric phase two-form~(\ref{eq:curv}) vanishes, as it should.

The case of $\eta=\pm 1$ is exceptional since it has degeneracies in the spectrum. In this case one may expect to find a $\mathrm{U}(2)$ non-Abelian geometric phase of the type discussed
by Wilczek and Zee \cite{wilczekzee}. However, we find that the phase is a diagonal subgroup of  
$\mathrm{U}(2)$ and  is essentially Abelian. This is easy to understand from time reversal symmetry.
Since translations and dilations are real operations, they commute with time reversal and so the allowed $\mathrm{U}(2)$ must also be real. This reduces $\mathrm{U}(2)$
to $\mathrm{O}(2)$, which is Abelian. By a suitable choice of  basis one can render the connection diagonal as in (86).
The ``non-Abelian'' $\mathrm{U}(2)$ Wilczek-Zee phase is in fact in an Abelian subgroup. 
It is also worth noting that the approach to $\eta=\pm 1$ is a singular limit because of the degeneracy there.

It is also interesting to note that the adiabatic transformations we consider act quite trivially on the spectrum of the Hamiltonian. Indeed, the translations are isospectral and the dilations
only cause an overall change in the scale of the energy spectrum $\lambda_n\rightarrow \lambda_n/l^2$. In particular, there are no level crossings and no degeneracies (away from 
$\eta\ne\pm 1$). This illustrates a remark made by Berry in the conclusion of \cite{berryphase}: although degeneracies play an important role in Berry's phase, they are not a necessary condition
for the existence of geometric phase factors. Indeed, our example reiterates this point. The Berry phases are nonzero even though one of the deformations is isospectral and the other a simple
scaling. It is the twisting of the eigenvectors over the parameter space that determines the Berry connection and phase, not the energy spectrum.

\ack 
This work was partially supported by PRIN 2010LLKJBX on ``Collective quantum phenomena: from strongly correlated systems to quantum simulators,'' and by the Italian National Group of Mathematical Physics (GNFM-INdAM).

\appendix

\section{Boundary triples}\label{appendix}
In this appendix we briefly recall the technique of boundary triples and their main applications to the search of self-adjoint extensions of densely defined symmetric operators. For a review on the subject see \cite{Brun}.

Von Neumann's theory of self-adjoint extensions does not provide an explicit way to construct them. The theorem, in fact, guarantees their existence once the dimensions of the deficiency subspaces are found to be equal.  However, self-adjoint extensions can be constructed as restrictions of the adjoint operator over suitable domains where a sesquilinear form identically vanishes. 

Given $T$ Hermitian, we define the following sesquilinear form:
\begin{eqnarray}
\label{eq:boundf}
& & \Gamma_{T^\dag}\,:\, \mathrm{D}(T^\dag)\times \mathrm{D}(T^\dag)\rightarrow \mathbb{C},\\
\nonumber
& &  \Gamma_{T^\dag}(\xi,\eta):=\langle T^\dag\xi | \eta\rangle-\langle\xi| T^\dag\eta\rangle, \qquad \forall \xi,\eta \in \mathrm{D}(T^\dag).
\end{eqnarray} 
The essential ingredient in the analysis of self-adjoint extensions is given by the deficiency subspaces, where the boundary form usually does not vanish.  Every element $\zeta\in\mathrm{D}(T^\dag)$ can be uniquely split into three components~\cite{Reed}
\begin{equation}
\zeta=\eta+\eta^{+}+\eta^{-},\qquad \eta \in \mathrm{D}(\overline{T}),\quad  \eta^{+}\in\mathrm{K}_+(T),\quad  \eta^{-}\in\mathrm{K}_-(T),
\end{equation}
where $\mathrm{K}_{\pm}(T)$ are the deficiency subspaces, that is the null spaces of $(T^\dag\mp \rmi\, \mathbb{I})$. From this decomposition it is easy to prove that
\begin{equation}
\label{eq:nonvan}
\Gamma_{T^\dag}(\zeta_1,\zeta_2)\,=\,2\rmi \left(\langle\eta_1^+ | \eta_2^+\rangle-\langle\eta_1^-|\eta_2^-\rangle\right),\qquad \forall \zeta_1,\zeta_2 \in\mathrm{D}(T^\dag)
\end{equation}
showing how the boundary form can be used as a measure of ``lack of self-adjointness".
Moreover von Neumann's theorem tells us that every self-adjoint extension is in a one-to-one correspondence with a unitary operator $\mathcal{U}$ between the deficiency subspaces. It follows that each self-adjoint extension of T is given by
\begin{equation}
\mathrm{D}(T_\mathcal{U})=\{\xi \in \mathrm{D}(T^\dag)\,:\,\Gamma_{T^\dag}(\xi,\eta_--\mathcal{U}\eta_-)=0\,,\,\forall \eta_-\in \mathrm{K}_-(T)\}.
 \end{equation}
 
Following \cite{Brun,Deoliv} we now introduce a more general tool useful for unveiling all the self-adjoint extensions of a symmetric operator. We will show how this naturally arises from von Neumann's theory and extends it. Moreover, von Neumann's theory and the use of boundary forms are helpful when studying differential operators, but what could one state about self-adjoint extensions of Hermitian operators, which are not in general differential operators? A possible answer could be given by boundary triples, which are a natural generalization of the notion of boundary values in functional spaces.

Let $T$ be a Hermitian operator with equal deficiency indices. Let \textit{h} be an auxiliary Hilbert space and take
\begin{equation}
\rho_1,\rho_2: \mathrm{D}(T^\dag)\rightarrow h ,
\end{equation}
which are supposed linear and with ranges  dense in \textit{h},
\begin{equation}
\overline{\mathrm{Ran} (\rho_1)}=\overline{\mathrm{Ran}(\rho_2)}=h.
\end{equation}
Suppose that they satisfy the following condition:
\begin{equation}
\langle\rho_1(\xi)|\rho_1(\eta)\rangle-\langle\rho_2(\xi)|\rho_2(\eta)\rangle = a\, \Gamma_{T^\dag}(\xi,\eta), \qquad \forall \xi,\eta \in \mathrm{D}(T^\dag), 
\end{equation}
where $a\in \mathbb{C}$, $a \ne 0$, and $\Gamma_{T^\dag}$ is the boundary form defined in~(\ref{eq:boundf}).
A triple $(h, \rho_1, \rho_2)$  that satisfies the above conditions is called a \textit{boundary triple}. 

Recall that from~(\ref{eq:nonvan}) the non-vanishing of the boundary form is due to non-trivial deficiency subspaces, so that one may choose either $h= \mathrm{K}_+(T)$ or $h=\mathrm{K}_-(T)$, and once more by von Neumann's theorem all self-adjoint extensions are in a one-to-one correspondence with unitary operators $\mathcal{U}:\mathrm{K}_-(T)\to \mathrm{K}_+(T)$.

Moreover, it could be useful to consider $h$ with the same dimension of either one of the two deficiency subspaces. The latter statement is enforced by the fact that two Hilbert spaces are unitarily equivalent if and only if they have the same dimension.

In general, it can be proved that given a boundary triple $(h,\rho_1,\rho_2)$ for a Hermitian operator with equal deficiency indices, all the self-adjoint extensions $T_\mathcal{U}$ of $T$ are given by
\begin{equation}
\mathrm{D}(T_\mathcal{U})=\left\{\xi \in \mathrm{D}(T^\dag)\,:\,\rho_2(\xi)=\mathcal{U}\rho_1(\xi) \right\},\qquad T_\mathcal{U}\xi=T^\dag\xi ,
\label{eq:exts}
\end{equation}
for every unitary operator $\mathcal{U}: h\to h$~\cite{Brun}.

In Section~\ref{BoundCond} we apply this theorem, by choosing as auxiliary space $h$ the space of boundary data, and a suitable pair of maps $\rho_1,\rho_2$, in order to get the 
parametrization of the self-adjoint extensions of the Laplacian exhibited in~\cite{Asorey1,Asorey3}.

\end{document}